\documentclass[conference, letterpaper]{IEEEtran}
\IEEEoverridecommandlockouts

\usepackage{cite}
\usepackage{amsmath,amssymb,amsfonts,bm}
\usepackage{textcomp}
\usepackage{xcolor}
\usepackage{scalerel}
\usepackage{mathtools}
\usepackage[T1]{fontenc}
\usepackage{siunitx}
\usepackage{tikz}
\usepackage{amsthm}
\usetikzlibrary{svg.path}
\usepackage[english]{babel}
\theoremstyle{definition}

\usepackage{esvect}

\usepackage{anyfontsize}
\usepackage{graphicx}
\usepackage{stfloats}
\usepackage{hyperref}
\usepackage{float}
\usepackage{multirow}

\usepackage{bbm}
\usepackage{subcaption}
\usepackage{mathrsfs}
\usepackage{pifont}
\usepackage{booktabs}
\usepackage{algorithm}
\usepackage{algpseudocode}

\usepackage[tableposition=top]{caption}

\definecolor{orcidlogocol}{HTML}{A6CE39}
\tikzset{
  orcidlogo/.pic={
    \fill[orcidlogocol] svg{M256,128c0,70.7-57.3,128-128,128C57.3,256,0,198.7,0,128C0,57.3,57.3,0,128,0C198.7,0,256,57.3,256,128z};
    \fill[white] svg{M86.3,186.2H70.9V79.1h15.4v48.4V186.2z}
                 svg{M108.9,79.1h41.6c39.6,0,57,28.3,57,53.6c0,27.5-21.5,53.6-56.8,53.6h-41.8V79.1z M124.3,172.4h24.5c34.9,0,42.9-26.5,42.9-39.7c0-21.5-13.7-39.7-43.7-39.7h-23.7V172.4z}
                 svg{M88.7,56.8c0,5.5-4.5,10.1-10.1,10.1c-5.6,0-10.1-4.6-10.1-10.1c0-5.6,4.5-10.1,10.1-10.1C84.2,46.7,88.7,51.3,88.7,56.8z};
  }
}
\newcommand\orcidicon[1]{\href{https://orcid.org/#1}{\mbox{\scalerel*{
\begin{tikzpicture}[yscale=-1,transform shape]
\pic{orcidlogo};
\end{tikzpicture}
}{|}}}}

\def\BibTeX{{\rm B\kern-.05em{\sc i\kern-.025em b}\kern-.08em
    T\kern-.1667em\lower.7ex\hbox{E}\kern-.125emX}}

\begin{document}



\title{RIS-Aided mmWave Localization Under Cross-Link Interference via Beam-Domain ML Fingerprinting}

\author{
    \IEEEauthorblockN{
        Md Tarek Hassan \orcidicon{0000-0002-9719-0204},
        Dmitry Zelenchuk \orcidicon{0000-0003-2866-629}, and
        Muhammad Ali Babar Abbasi \orcidicon{0000-0002-1283-4614}
    }
    \IEEEauthorblockA{
        Centre for Wireless Innovation (CWI), School of Electronics,
        Electrical Engineering and Computer Science (EEECS),\\
        Queen's University Belfast, U.K.
        Emails: \{mhassan15, d.zelenchuk, m.abbasi\}@qub.ac.uk.
    }
    \thanks{
    
    This work was supported in part by the Department for the Economy of Northern Ireland under the U.S. Ireland Research and Development Partnership under Grant USI 269, in part by the Horizon Europe Project MULTIPLY-6G and in part by the Project REASON (Research on Enabling Architectures and Solutions for Open Networks) sponsored by the Department of Science, Innovation and Technology (DSIT). M.\,T.\,Hassan is supported by EPSRC and is also with Rajshahi University of Engineering \& Technology (RUET), Bangladesh.
    }
}

\maketitle

\begin{abstract}
Accurate user equipment (UE) localization is critical for beam management in reconfigurable intelligent surface (RIS)-assisted millimeter-wave (mmWave) based sixth-generation (6G) networks, especially if the direct base-station--UE links are unavailable. This paper proposes a beam-domain fingerprint framework that maps the received signal-to-noise ratio (SNR) across a small set of predefined RIS reflection states to the UE azimuth angle and range, without requiring channel state information (CSI). Crucially, we extend the framework to a realistic interference-impaired scenario in which a nearby cross-link interferer (CLI) corrupts the clean SNR fingerprint, yielding a signal-to-interference-plus-noise 
ratio (SINR) fingerprint; an interference-to-noise ratio (INR)-constrained calibration strategy keeps the interference level physically interpretable. Four machine-learning (ML) regressors are evaluated under both conditions. Simulation results at 28\,GHz with a $20\!\times\!20$ RIS show that k-nearest neighbors (KNN) achieves the lowest angle MAE of $0.37$ degrees and range MAE of $4$ cm under clean conditions, rising to $1.4$ degrees and $7.6$ cm under interference. A key finding is that interference degrades angle estimation substantially more than range estimation across all models, a consequence of the asymmetric encoding of location information in the beam-domain fingerprint.
\end{abstract}


\section{Introduction}
\label{sec:intro}

Sixth-generation~(6G) wireless networks are expected to support applications such as indoor navigation, autonomous robotics, and extended reality that demand sub-meter positioning accuracy as a native service rather than an add-on~\cite{tataria20216g}. Millimeter-wave~(mmWave) bands are naturally aligned with this requirement: their short wavelengths produce fine angular resolution, and their multi-hundred-megahertz bandwidths support centimeter-level ranging in principle~\cite{shahmansoori2018position}. However, mmWave propagation is inherently fragile, as buildings, vehicles, and even human bodies can obstruct the direct path entirely, rendering the link unusable \cite{elmossallamy2020ris}. Reconfigurable intelligent surfaces~(RISs) have emerged as a practical remedy by acting as programmable phase-shifting mirrors that create virtual line-of-sight paths around obstacles, restoring both communication and sensing coverage in otherwise blocked environments~\cite{9837936}.
Beyond improving coverage, RISs offer a structural advantage for localization that is absent in conventional deployments, enabling users to be within the array's near-field region. Hence, the received wavefront exhibits measurable curvature that varies distinctly from one spatial point to another~\cite{dardari2022losnlos,emenonye2024singleanchor}, encoding range information directly into the received signal and enabling joint azimuth and distance estimation from a single RIS with no additional anchor~\cite{lu2024nearfieldxlmimo}. For mmWave frequencies, the Rayleigh distance of even a moderately sized RIS falls within the typical indoor operating range, placing users firmly in the near field and making this effect practically relevant~\cite{liu2025nearfieldsurvey}.

Model-based estimators can exploit near-field wavefront curvature through maximum-likelihood or subspace methods, but they rely on a precisely known channel model~\cite{elzanaty2021risloc}. In practice, however, 1-bit phase quantization adopted in established low-cost, low-complexity RIS architectures combined with a discrete beam codebook and the strongly nonlinear mapping that arises when the RIS steers into a limited number of fixed directions, substantially degrades the accuracy of analytic estimators~\cite{lin2022channel}. Moreover, such methods require explicit channel state information~(CSI) estimation, whose overhead scales with the array size and becomes prohibitive for large apertures~\cite{lu2024nearfieldxlmimo}. These limitations motivate a data-driven alternative. Specifically, rather than deriving position analytically, a machine-learning~(ML) regression model learns to map the received measurements to location directly from labeled training data samples~\cite{katla2019deep,wang2021jointbeam}. A compact and attractive measurement vector for this purpose is a beam-domain SNR fingerprint, formed by recording the received signal-to-noise ratio~(SNR) under each of the predefined RIS reflection states. Each state illuminates the environment from a distinct angular direction, so the resulting dimensional vector encodes the user's spatial signature without any explicit CSI  estimation~\cite{wang2021jointbeam,li2023jointbeamforming}. In a fully blocked base station (BS)-user equipment (UE) scenario, where the RIS provides the only communication path as illustrated in Fig.~\ref{fig:Scenario}, this cascaded fingerprint is the only available localization cue, yet the probing overhead with small probing beam states remains modest.

Despite this promise, one practical complication has received very little attention from a localization perspective. In a multi-user mmWave deployment under flexible time-division duplex (TDD) scheduling, a UE transmitting uplink in one cell can cause cross-link interference (CLI) in a neighboring cell operating on the same mmWave carrier frequency~\cite{3gpp38828}.
In RIS-assisted deployments, CLI is further compounded because the passive RIS reflects all impinging signals without discrimination, scattering the CLI signal toward the target UE in a beam-state-dependent manner and converting the clean SNR fingerprint into a 
signal-to-interference-plus-noise-ratio~(SINR) fingerprint at inference time. Since the ML model is trained on interference-free data, this mismatch between training and inference directly degrades localization accuracy. While prior work has addressed RIS-assisted localization~\cite{wang2021jointbeam}, joint beamforming and positioning~\cite{li2023jointbeamforming}, and 
multi-user RIS systems~\cite{zhang2021metalocalization}, the impact of CLI-induced SINR fingerprint corruption on ML-based near-field joint angle and range estimation is largely unexplored, an important gap this paper fills. The main contributions of this paper are as follows.
\begin{itemize}

\item A near-field focused BS--RIS--UE localization framework is proposed using a compact $B$-dimensional beam-domain SNR fingerprint as the ML input for joint azimuth-angle and range estimation.

\item A CLI model is introduced where a nearby UE in an adjacent cell induces beam-state-dependent leakage via RIS sidelobe scattering, corrupting the clean SNR fingerprint into a SINR fingerprint at inference time; an INR-constrained calibration ensures physical interpretability.

\item SINR corruption degrades angle estimation far more severely than range under CLI among the four ML regressor models evaluated.

\end{itemize}

The remainder of the paper is organized as follows.
Section~\ref{System_model} describes the system and channel models. Section~\ref{sec_iii} presents the ML-based localization framework and the interference model. Section~\ref{RDSection} evaluates performance under clean and interfered conditions, and Section~\ref{Conclusion} concludes.

\section{RIS-Assisted System Model}
\label{System_model}

\subsection{System Geometry}

\begin{figure}[!htbp]
    \centering
    \includegraphics[width=\linewidth]{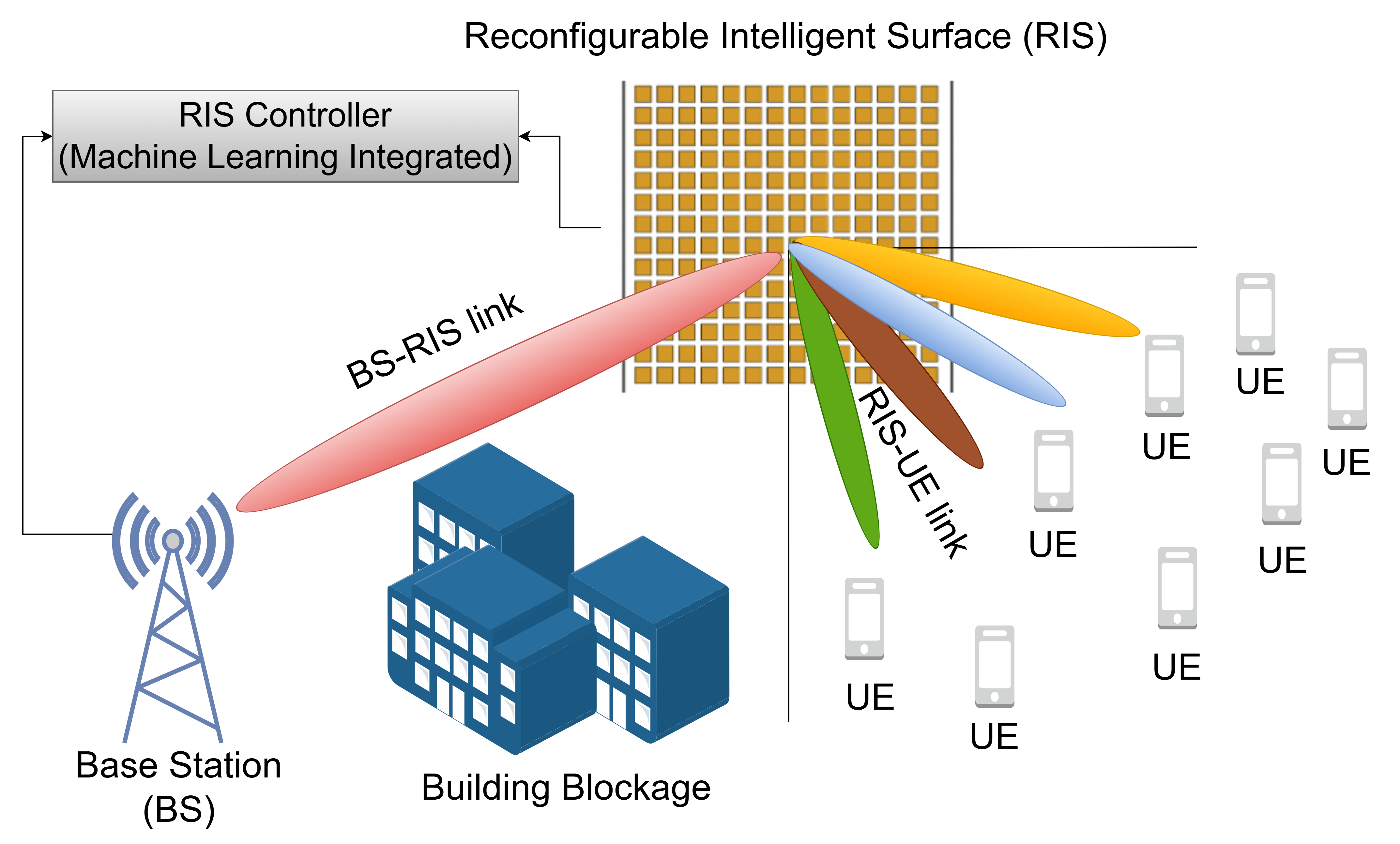}
    \caption{RIS-aided localization system model.}
    \label{fig:Scenario}
    \vspace{-4 mm}
\end{figure}

We consider an RIS-assisted mmWave localization system operating at a carrier frequency of $f=28$\,GHz, where the direct BS--UE link is blocked, as illustrated in Fig.~\ref{fig:Scenario}. The study is based on a down-scaled experimental scenario using a $20 \times 20$ 1-bit RIS developed in our laboratory at Queen's University Belfast~\cite{Gabriel}. For practical pico-cell or macro-cell deployments, both the RIS aperture and the localization region can be scaled accordingly. The RIS is deployed on the $yz$-plane and consists of $L=M_yN_z$ passive elements arranged as a uniform planar array, with inter-element spacings $d_y$ and $d_z$.
The BS is fixed at $\mathbf{p}_{\mathrm{BS}}=(x_t,y_t,z_t)$ and the target UE at $\mathbf{p}_u=(x_u,y_u,z_u)$. With a fixed UE elevation angle $\theta_r=90^\circ$, the position reduces to two unknowns: azimuth $\varphi_u$ from RIS to UE and range $d_u$ from the RIS center, with $x_u=d_u\cos\varphi_u$ and $y_u=d_u\sin\varphi_u$. All BS and UE distances satisfy the near-field condition $d < 2D^2/\lambda$ (where $D$ is the effective aperture dimension) for the $20\!\times\!20$ RIS at 28\, GHz as illustrated in Fig.~\ref{fig:inference}(a), so the wavefront is spherical and the received signal carries both azimuth and range information.

\subsection{Element-Wise Pattern Model}
Let $r_{t,l}=\|\mathbf{p}_{\mathrm{BS}}-\mathbf{p}_l\|$, $r_{u,l}=\|\mathbf{p}_u-\mathbf{p}_l\|$, and
$\rho_l=\sqrt{y_l^2+z_l^2}$ be the BS-to-element, element-to-UE, and element-to-RIS-center distances, respectively. Each element is modeled as a cosine-pattern radiator. Applying the law of cosines to the triangle formed by the BS, the RIS centre, and element $l$, the BS-side off-boresight angle
satisfies
\begin{equation}
\cos\alpha_{t,l}
= \frac{d_1^2+r_{t,l}^2-\rho_l^2}{2\,d_1\,r_{t,l}},
\label{eq:cos_alpha_t}
\end{equation}
where $d_1=\|\mathbf{p}_{\mathrm{BS}}\|$ is the BS--RIS distance. The incidence projection onto the RIS surface is
\begin{equation}
\cos\theta_{t,l}^{\mathrm{inc}} = \frac{x_t}{r_{t,l}}.
\label{eq:cos_theta_t}
\end{equation}
Combining both, the BS-side power pattern factor is
\begin{equation}
F_{t,l}
= \bigl(\cos\alpha_{t,l}\bigr)^{q_t}
  \cos\theta_{t,l}^{\mathrm{inc}},
\label{eq:F_t}
\end{equation}
where $q_t = G_t/2-1$ is the pattern exponent that maps transmit antenna gain $G_t$ (linear) with respect to a cosine radiation model. The analogous UE-side quantities are \cite{9837936}
\begin{equation}
\cos\alpha_{u,l}
= \frac{d_u^2+r_{u,l}^2-\rho_l^2}{2\,d_u\,r_{u,l}},
\qquad
\cos\theta_{u,l}^{\mathrm{ref}} = \frac{x_u}{r_{u,l}},
\label{eq:cos_u}
\end{equation}
\begin{equation}
F_{u,l}
= \bigl(\cos\alpha_{u,l}\bigr)^{q_r}
  \cos\theta_{u,l}^{\mathrm{ref}},
\label{eq:F_u}
\end{equation}
where $q_r = G_r/2-1$ and $G_r$ is the receive antenna gain.

\subsection{Cascaded BS--RIS--UE Channel Model}

The RIS applies $B$ predefined reflection states sequentially; index $b\in\{1,\dots,B\}$ identifies the active state. Under state $b$, element $l$ is assigned the 1-bit phase shift $\phi_l^{(b)}\in\{0,\pi\}$ determined by the codebook steering
direction $\phi_b$:
\begin{equation}
\phi_l^{(b)}
= \begin{cases}
    0,    & \phi_{\mathrm{pp},l}(\phi_b) \bmod 2\pi \in [0,\pi), \\
    \pi,  & \text{otherwise,}
  \end{cases}
\label{eq:1bit}
\end{equation}
where $\phi_{\mathrm{pp},l}(\phi_b)=-k_0 y_l\sin\phi_b$ is the ideal progressive phase at element $l$ for azimuth $\phi_b$, and $k_0=2\pi/\lambda$ is the free-space wavenumber. The reflection coefficient is $\Gamma_l^{(b)}=\beta\,e^{j\phi_l^{(b)}}$, where $\beta\in(0,1]$ is the element reflection amplitude. The effective cascaded channel from the BS to user $u$ is
\begin{equation}
h_u^{(b)}
= \frac{A_e}{4\pi}
  \sum_{l=1}^{L}
  \frac{\sqrt{F_{t,l}\,F_{u,l}}}{r_{t,l}\,r_{u,l}}
  \,\Gamma_l^{(b)}\,
  \exp\!\bigl[-jk_0(r_{t,l}+r_{u,l})\bigr],
\label{eq:heff}
\end{equation}
where $A_e=d_y d_z$ is the element area. The power-domain factors $F_{t,l}$ and $F_{u,l}$ appear under
square roots in~\eqref{eq:heff} because they are power patterns, whereas the channel coefficient is a field quantity. The received power at user $u$ is
\begin{equation}
P_u^{(b)} = P_t G_t G_r \bigl|h_u^{(b)}\bigr|^2,
\label{eq:rx_power}
\end{equation}
where $P_t$ is the BS transmit power.

\subsection{Beam-Domain SNR Fingerprint}

The deterministic thermal noise power in this scenario is
$\sigma^2 = k_B T_0 B_w F_n \approx -80.95$\,dBm,
where $k_B$ is Boltzmann's constant, $T_0=290$\,K, $B_w=400$\,MHz, and $F_n=7$\,dB is the receiver noise figure. The per-state SNR at user $u$ is
\begin{equation}
\gamma_u^{(b)} = \frac{P_u^{(b)}}{\sigma^2}.
\label{eq:snr}
\end{equation}
Activating all $B$ states yields the beam-domain fingerprint
\begin{equation}
\mathbf{x}_u =
\bigl[\gamma_u^{(1)},\,\gamma_u^{(2)},\,\dots,\,\gamma_u^{(B)}\bigr]^{\mathsf T}\!,
\label{eq:fingerprint}
\end{equation}
which encodes the user's spatial signature without requiring CSI. This fingerprint is passed to a trained regression model,
\begin{equation}
[\hat\varphi_u,\,\hat d_u] = f_{\mathrm{ML}}(\mathbf{x}_u),
\label{eq:ml}
\end{equation}
where $f_{\mathrm{ML}}(\cdot)$ maps the $B$-dimensional observation to the two localization parameters.

\section{Proposed Interference-based Localization Approach}
\label{sec_iii}
\subsection{Interference-Aware Inference Model}
\label{subsec:interference}
\begin{figure*}[!htbp]
    \centering
    \begin{subfigure}[b]{0.37\linewidth}
    \includegraphics[width=\linewidth]{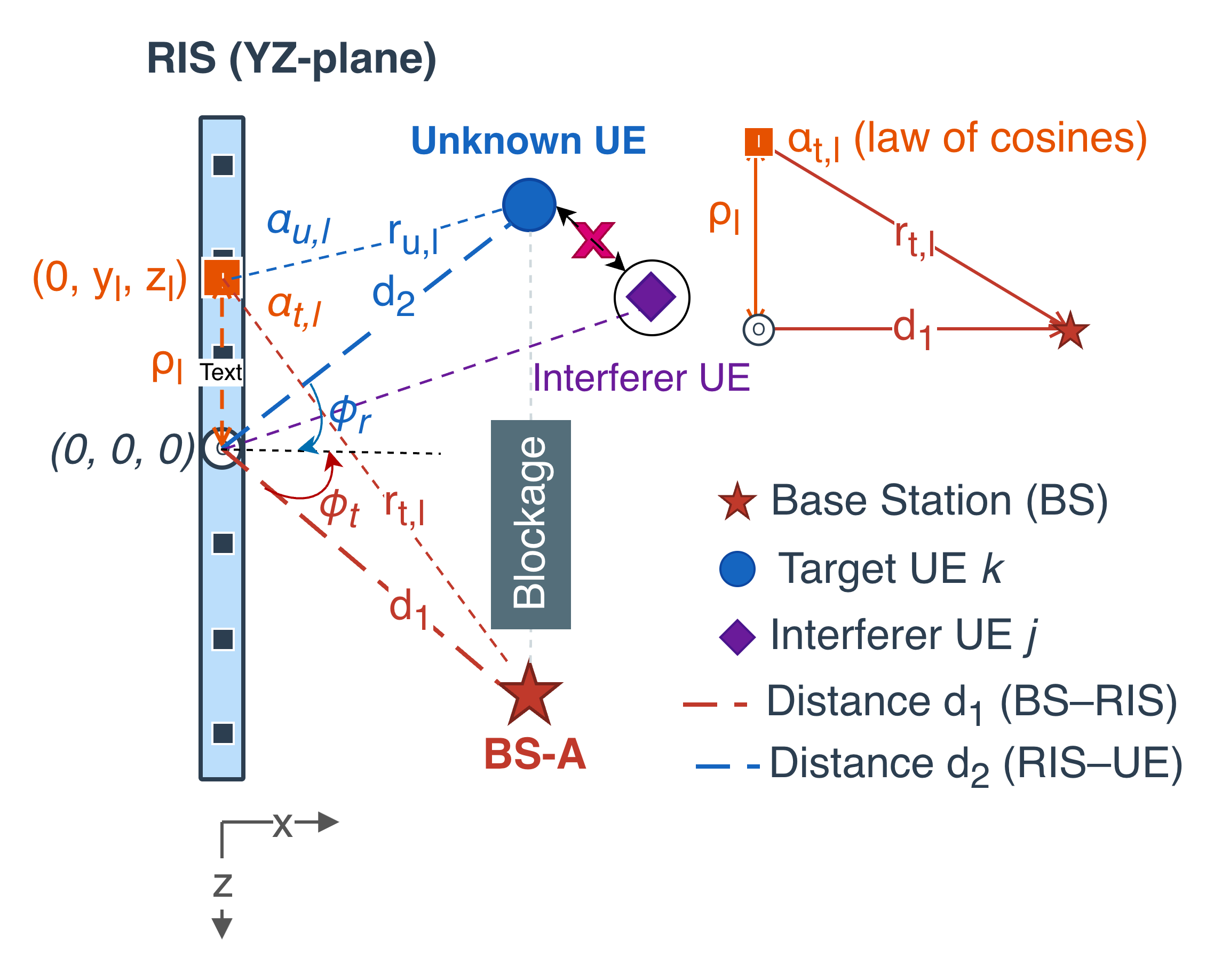}
    \caption{Near-field RIS geometry.}
    \label{fig:inference_geometry}
    \end{subfigure}
    \hfill
    \begin{subfigure}[b]{0.60\linewidth}
    \includegraphics[width=\linewidth]{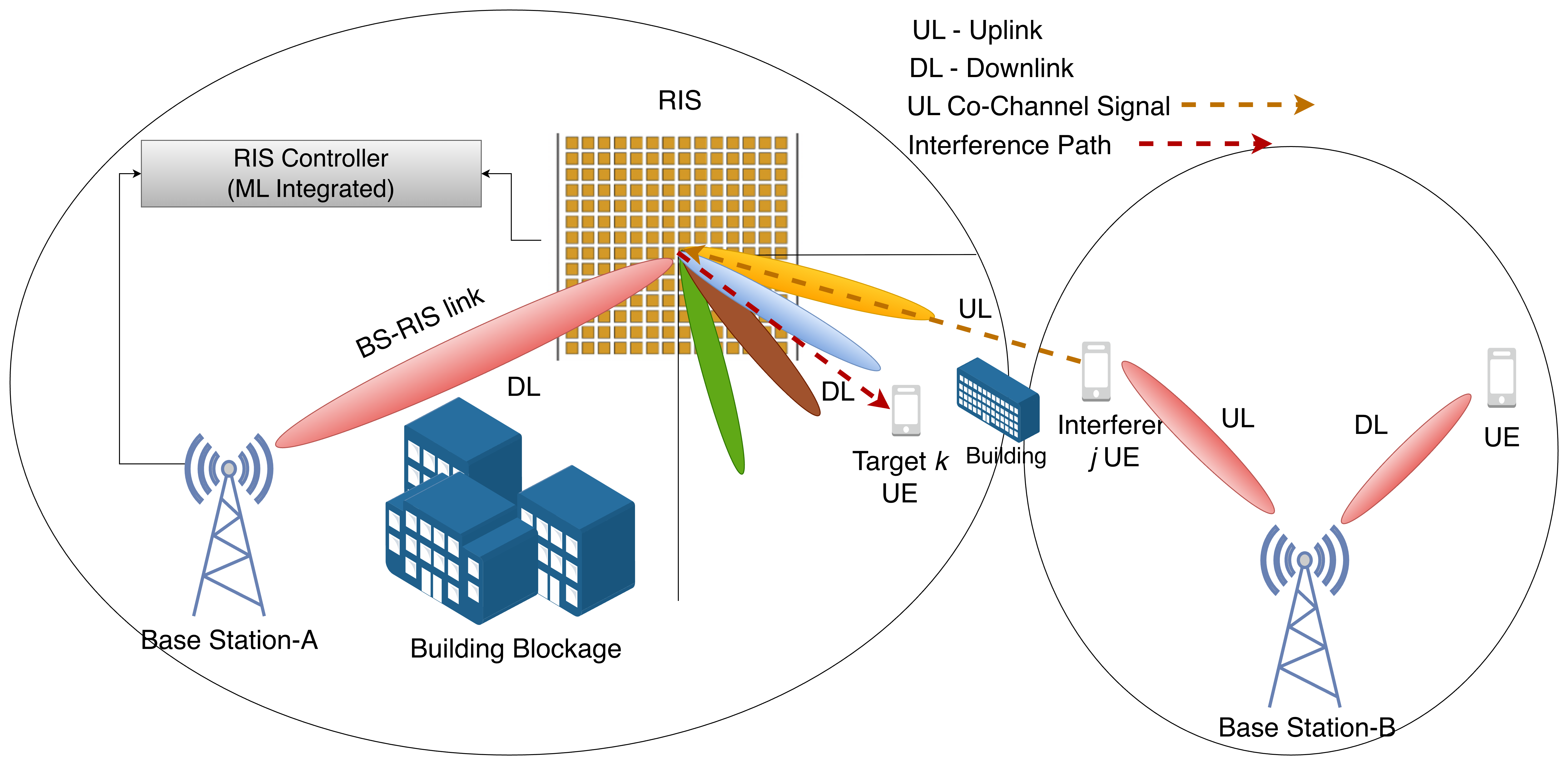}
    \caption{Two-cell deployment scenario showing CLI.}
    \label{fig:inference_scenario}
    \end{subfigure}
    \caption{System model. (a) Near-field geometry of the RIS-aided BS--UE link. (b) A cross-link interferer UE$(j)$ in an adjacent cell transmits uplink on the same carrier frequency; its signal impinges on the passive RIS of BS-A and scatters toward the target UE$(k)$, corrupting the SNR fingerprint into a SINR fingerprint.}
    \vspace{-5mm}
    \label{fig:inference}
\end{figure*}
At inference time, a cross-link interferer UE$(j)$ from an adjacent BS-B transmits uplink on the same carrier frequency as the localization pilot of BS-A in Fig.~\ref{fig:inference}(b). Because both BSs operate on the same mmWave carrier frequency, UE$(j)$'s uplink signal propagates through the shared medium and impinges on the passive RIS surface of BS-A. The RIS cannot discriminate between the intended downlink pilot from BS-A and the cross-link uplink signal from UE$(j)$: both are reflected according to the active phase profile~$\Phi_b$. A fraction of UE$(j)$'s uplink energy therefore scatters toward UE$(k)$ as beam-state-dependent interference, whose strength is governed by the sidelobe coupling $\omega_j^{(b)}$ of the active beam profile toward UE$(j)$'s angular direction. 
\color{black}
The normalized leakage weight is expressed as
\begin{equation}
\omega_j^{(b)} =
\frac{|h_j^{(b)}|^2}{\max_{m}\,|h_j^{(m)}|^2 + \varepsilon},
\quad \omega_j^{(b)}\in[0,1],
\label{eq:omega}
\end{equation}
which captures the relative RIS coupling gain toward user~$j$ 
under beam state~$b$; since beam~$b$ is optimized for 
user~$k$, coupling to user~$j$ arises through sidelobe 
leakage and $\omega_j^{(b)}\ll 1$ in general. The beam-dependent interference coefficient is
\begin{equation}
\xi_b = \eta_{\mathrm{eff}}\,\rho\,\frac{P_j}{P_k}\,\omega_j^{(b)},
\label{eq:xi}
\end{equation}
where $P_j/P_k$ is the cross-link power ratio, $\rho\in(0,1]$ is the temporal overlap factor, which is fraction of the probing interval during which user~$j$ transmits simultaneously with user~$k$'s pilot reception, and $\eta_{\mathrm{eff}}\in(0,1]$ is the effective coupling factor calibrated below. The per-beam INR-based SINR at user~$ k$ is then
\begin{equation}
\mathrm{INR}_k^{(b)} = \xi_b\,\gamma_k^{(b)},
\label{eq:inr}
\end{equation}
\begin{equation}
\tilde{\gamma}_k^{(b)}
= \frac{\gamma_k^{(b)}}{1 + \xi_b\,\gamma_k^{(b)}}.
\label{eq:sinr}
\end{equation}
\begin{algorithm}[!tb]
\caption{Beam-Domain Fingerprint Generation and 
         Interference-Aware SINR Localization}
\label{alg:combined}
\begin{algorithmic}[1]
\Require RIS geometry, BS/interferer positions, azimuth angle set $\mathcal{A}_\varphi$,
         $K$ distances, $B$ reflection states, system parameters,
         test set $\mathcal{S}$, regressors $\{f_m\}$
\Ensure SNR fingerprint dataset $\mathcal{D}$; 
        SINR-based localization
\State \textbf{--- Phase 1: Training Fingerprint Generation ---}
\State Pre-compute BS-side distances and pattern factors
\ForAll{$\varphi_u\in\mathcal{A}_\varphi  $}
  \State Draw $K$ distances $\{d_u^{(k)}\}_{k=1}^{K}$
  \For{$k=1$ to $K$}
    \For{$b=1$ to $B$}
      \State Compute $h_u^{(b)}$, $P_u^{(b)}$, $\gamma_u^{(b)}$
             via \eqref{eq:heff}--\eqref{eq:snr}
    \EndFor
    \State Form $\mathbf{x}_u$ via~\eqref{eq:fingerprint};
           store $(\mathbf{x}_u,\varphi_u,d_u^{(k)})$ in $\mathcal{D}$
  \EndFor
\EndFor
\State Train regressors $\{f_m\}$ on $\mathcal{D}$
\State \textbf{--- Phase 2: Interference-Aware Inference ---}
\State Compute $\{h_j^{(b)}\}$, $\{\omega_j^{(b)}\}$ 
       via \eqref{eq:heff}, \eqref{eq:omega}
\ForAll{$i\in\mathcal{S}$}
  \For{$b=1$ to $B$}
    \State Compute $\mathrm{INR}_i^{(b)}$, 
           $\tilde\gamma_i^{(b)}$ via \eqref{eq:inr}--\eqref{eq:sinr}
  \EndFor
  \State Form $\tilde{\mathbf{x}}_i$; 
         predict via \eqref{eq:ml} and \eqref{eq:ml_sinr}
\EndFor
\State \Return $\mathcal{D}$;\; $\hat\varphi_k^{\mathrm{SINR}},\,\hat{d}_k^{\mathrm{SINR}}$
\end{algorithmic}
\end{algorithm}
To prevent unrealistically severe degradation,
$\eta_{\mathrm{eff}}$ is chosen so that the maximum INR over all
test samples $i\in\mathcal{S}$ and beam states does not exceed
$\Gamma_{\max}^{\mathrm{lin}}=10$\,dB.
The pre-calibration coupling term is defined as
\begin{equation}
\zeta_{i,b}
= \rho\,\frac{P_j}{P_k}\,\omega_j^{(b)}\,\gamma_i^{(b)},
\quad i\in\mathcal{S},
\label{eq:zeta}
\end{equation}
Based on that, the coupling factor is set to
\begin{equation}
\eta_{\mathrm{eff}}
= \min\!\left(\eta,\;
  \frac{\Gamma_{\max}^{\mathrm{lin}}}{\max_{i,b}\,\zeta_{i,b}}\right),
\label{eq:eta_eff}
\end{equation}
guaranteeing $\max_{i,b}\,\mathrm{INR}_k^{(b)}
\le\Gamma_{\max}^{\mathrm{lin}}$.
The corrupted fingerprint
$\tilde{\mathbf{x}}_k =
[\tilde{\gamma}_k^{(1)},\dots,\tilde{\gamma}_k^{(B)}]^{\mathsf{T}}$
is passed to the same trained model as in~\eqref{eq:ml}
\begin{equation}
[\hat{\varphi}_k^{\mathrm{SINR}},\,\hat{d}_k^{\mathrm{SINR}}]
= f_{\mathrm{ML}}\!\left(\tilde{\mathbf{x}}_k\right).
\label{eq:ml_sinr}
\end{equation}
Since $f_{\mathrm{ML}}$ is trained on clean fingerprints only, any accuracy loss directly quantifies each regressor's robustness to fingerprint corruption without retraining.
\begin{table*}[!t]
\centering
\caption{Comparison of SNR-based and Interference-Aware
SINR-based localization performance.}
\label{tab:performance}
\setlength{\tabcolsep}{6pt}
\renewcommand{\arraystretch}{1.12}
\begin{tabular}{l|cc|cc|cc|cc|cccc}
\toprule
\multirow{2}{*}{\textbf{Model}}
  & \multicolumn{2}{c|}{\textbf{SNR Angle}}
  & \multicolumn{2}{c|}{\textbf{SNR Range}}
  & \multicolumn{2}{c|}{\textbf{SINR Angle}}
  & \multicolumn{2}{c|}{\textbf{SINR Range}}
  & \multicolumn{4}{c}{\textbf{Performance Change}} \\
\cmidrule(lr){2-3}\cmidrule(lr){4-5}\cmidrule(lr){6-7}
\cmidrule(lr){8-9}\cmidrule(lr){10-13}
  & MAE ($^\circ$) & $R^2$
  & MAE (m) & $R^2$
  & MAE ($^\circ$) & $R^2$
  & MAE (m) & $R^2$
  & $\Delta\phi_r$ ($^\circ$) & $\Delta R^2$
  & $\Delta r$ (m) & $\Delta R^2$ \\
\midrule
\textbf{KNN} & \textbf{0.370} & \textbf{0.997} & \textbf{0.040} & \textbf{0.958}
             & \textbf{1.392} & \textbf{0.967} & \textbf{0.076} & \textbf{0.900}
             & $+1.022$ & $-0.030$ & $+0.036$ & $-0.058$ \\
RF  & 3.223 & 0.909 & 0.054 & 0.942
             & 4.607 & 0.871 & 0.083 & 0.877
             & $+1.384$ & $-0.038$ & $+0.029$ & $-0.065$ \\
DT  & 3.992 & 0.742 & 0.060 & 0.910
             & 5.833 & 0.661 & 0.105 & 0.751
             & $+1.841$ & $-0.081$ & $+0.045$ & $-0.159$ \\
SVR & 8.125 & 0.626 & 0.092 & 0.872
             & 9.837 & 0.581 & 0.116 & 0.804
             & $+1.712$ & $-0.045$ & $+0.024$ & $-0.068$ \\
\bottomrule
\end{tabular}
\end{table*}
\subsection{Dataset, Regression Models and RIS Phase Generation}
\label{subsec:dataset}
The training dataset $\mathcal{D}$ comprises $N=600$ labeled examples, built by sweeping UE azimuth over 30 uniformly-spaced angles with $K=20$ distances per angle. For each $(\varphi_u,d_u)$, four 1-bit beam states---sufficient for angular discrimination 
across $0^\circ$--$90^\circ$---are probed and the SNR fingerprint formed via~\eqref{eq:fingerprint} in algorithm \ref{alg:combined}. Classical regressors, such as k-nearest neighbors (KNN), random forests (RF), decision trees (DT), and support vector regressors (SVR) are chosen over deep learning to isolate CLI effects without confounding factors such as architecture or data volume. All are trained on an 80/20 split of clean fingerprints and tested under both SNR and SINR conditions with the same feature scaler. Given $\hat\varphi_u$, the RIS controller selects the nearest codebook beam and steers toward the unknown UE.
\section{Simulation Results and Discussion}
\label{RDSection}
\subsection{Simulation Settings and ML Performance Parameters}
We consider a $20\times20$ RIS with $d_y=d_z=4.6$\, mm and $\beta=0.7$. The BS is placed at $d_1=0.8$\,m with $(\theta_t,\varphi_t)=(90^\circ,-30^\circ)$, while the UE elevation is fixed at $\theta_r=90^\circ$ and the azimuth varies within $0^\circ\leq\varphi_r\leq90^\circ$. The transmit power is $P_t=20$\,dBm, and the antenna gains are $G_t=15$\,dBi and $G_r=5$\,dBi. Four beam SNR measurements at each UE point for four swept beams \(\phi_b \in \{11^\circ,33^\circ,56^\circ,78^\circ\}\) are used as the ML inputs, whereas the outputs are the UE azimuth $\varphi_r$ and range $d_2$. For the interference case, a nearby user at $\varphi_j=45^\circ$ and $d_{2,j}=0.48$\,m is introduced only at inference, with $P_j/P_k=0.25$, $\eta=0.20$, $\rho=1.00$, and an INR cap of 10\,dB.
\begin{figure*}[!tb]
    \centering
    \subfloat[]{%
        \includegraphics[width=0.62\textwidth]{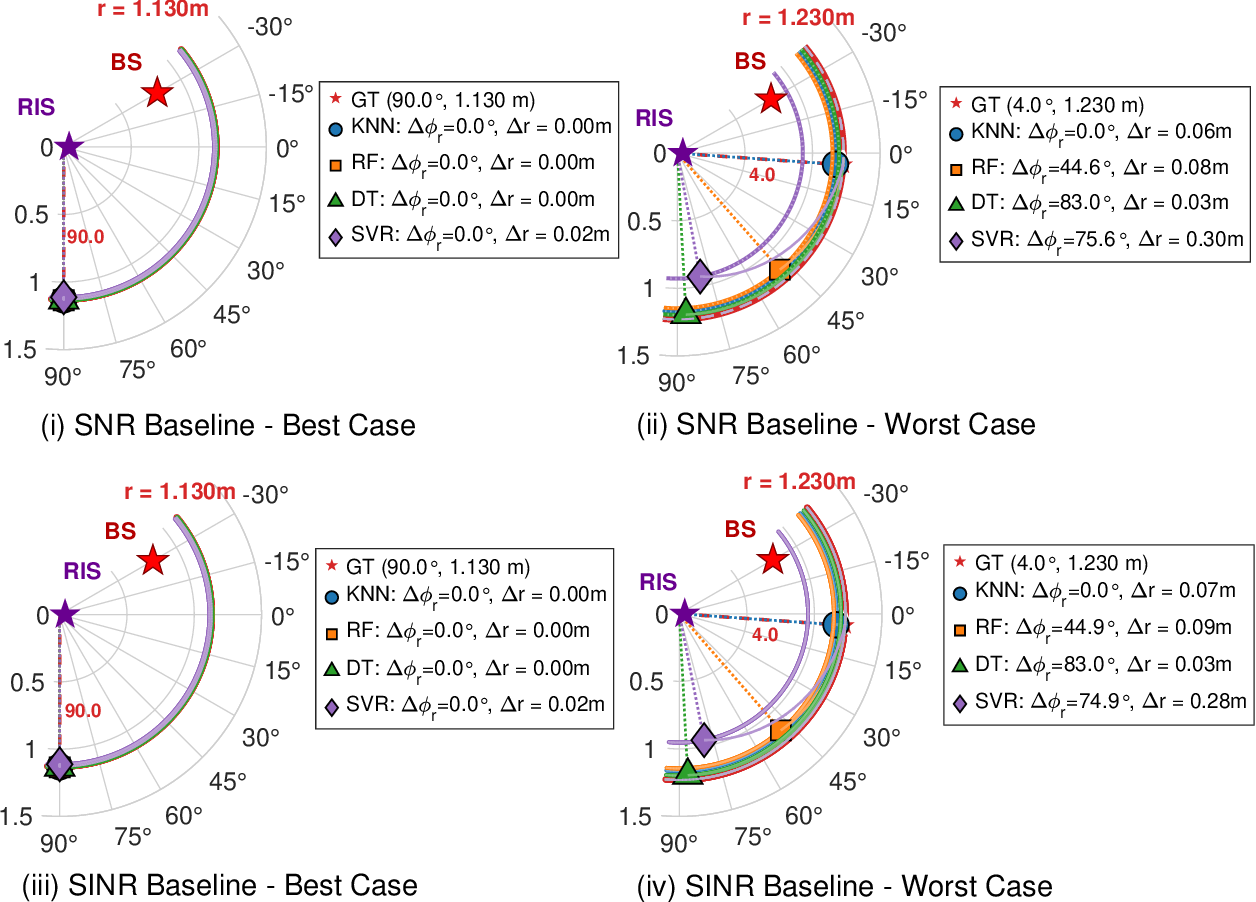}%
        \label{fig:polar}
    }
    \hfill
    \subfloat[]{%
        \includegraphics[width=0.35\textwidth]{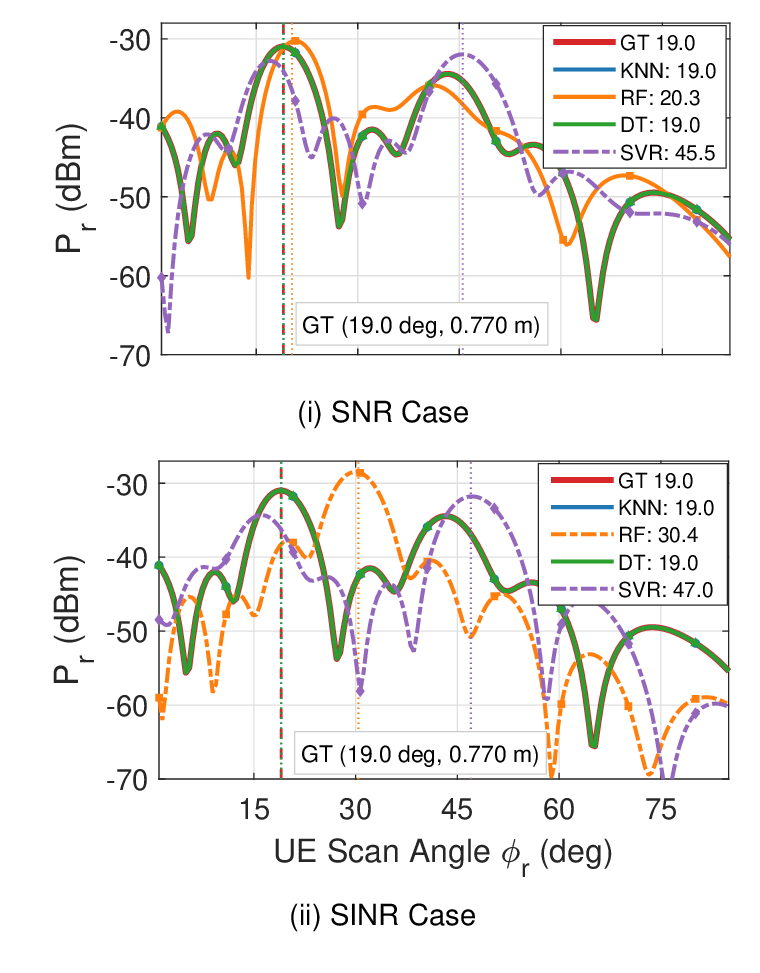}%
        \label{fig:radiation}
    }
    \caption{(a) Polar localization results: SNR and SINR best-worst case. 
    (b) RIS radiation pattern for GT $(19^\circ,\,0.770\,\mathrm{m})$ under SNR and SINR scenarios.}
    \label{fig:combined_side}
      \vspace{-2 mm}
\end{figure*}
\begin{figure}[!htbp]
  \centering
  \includegraphics[width=0.92\columnwidth]{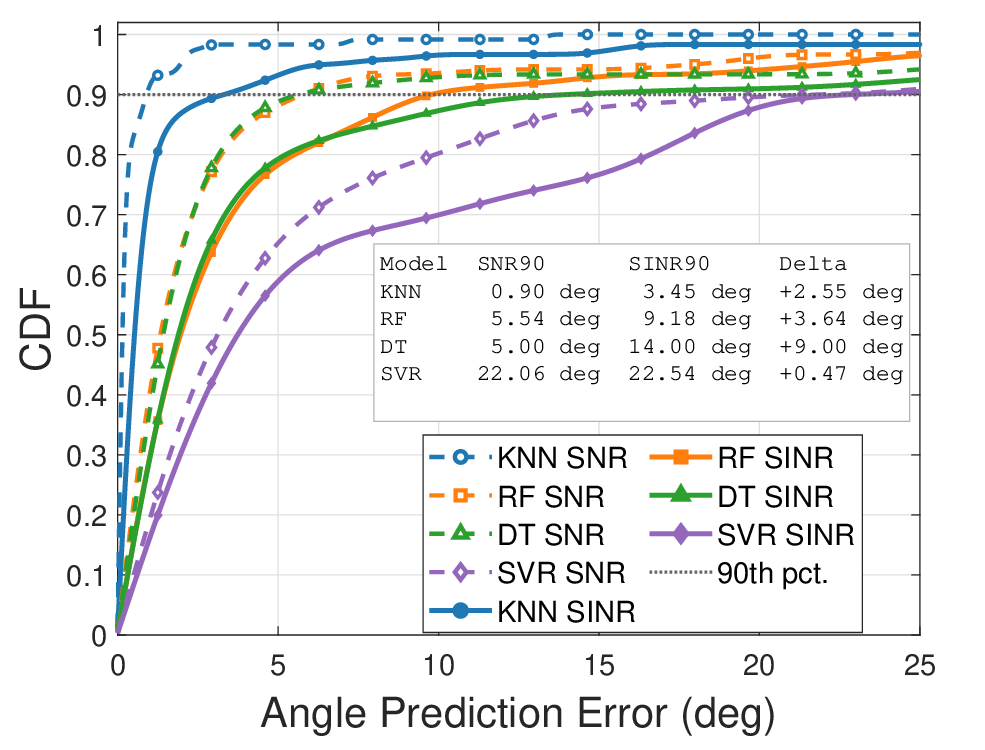}
  \caption{CDF of angle error (dashed\,=\,SNR, solid\,=\,SINR).}
  \label{fig:cdf_angle}
  \vspace{-4 mm}
\end{figure}
\noindent Mean absolute error (MAE) measures the average absolute difference between predicted and true values; lower is better. The coefficient of determination $R^2$ measures how well predictions track the ground-truth trend, with $R^2\!=\!1$ indicating a perfect fit.
\begin{figure}[!htbp]
  \centering
  \includegraphics[width=0.92\columnwidth]{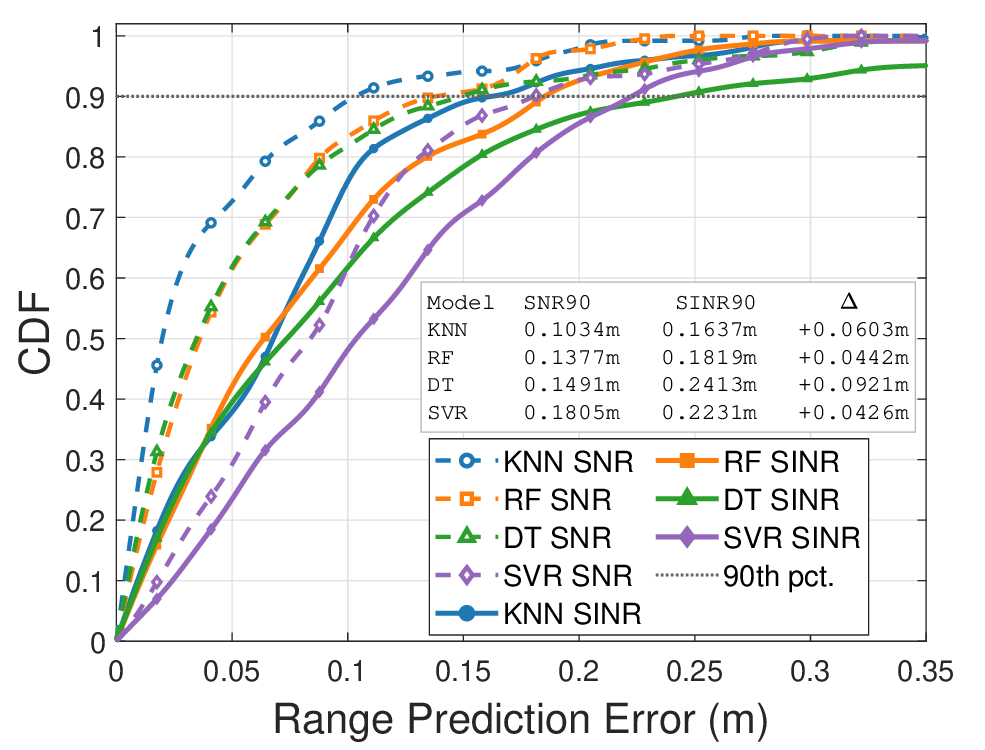}
  \caption{CDF of range error (dashed\,=\,SNR, solid\,=\,SINR).}
  \label{fig:cdf_range}
    \vspace{-4 mm}
\end{figure}
Table~\ref{tab:performance} compares prediction accuracy under the SNR baseline and the SINR interference scenario. Under clean conditions, KNN achieves the best angle MAE of $0.37^\circ$ and range MAE of $4$\, cm, outperforming RF, DT, and SVR by a large margin. Under CLI, all models degrade, yet KNN remains the most robust, with angle and range MAE increases of only $+1.022^\circ$ and $3.6$\, cm, respectively. A central observation is that $\Delta\phi_r$ consistently exceeds $\Delta r$ across all models: interference impairs angle estimation far more than range estimation, a finding elaborated in Section~\ref{subsec:cdfs}.
\subsection{Spatial Localization Validation}
Fig.~\ref{fig:combined_side}(a) shows polar-domain predictions for the best and worst localization cases. Each subplot overlays the GT  arc (red), predicted-range arcs, and predicted-angle spokes, where the annular band visualizes range error $\Delta r$ and spoke offset shows angle error $\Delta\phi$. In the best case (i,\, iii), all models cluster tightly around the GT under SNR with only marginal drift under SINR. In the worst case (ii,\,iv), the GT lies at $4^\circ$ near the beam-coverage edge, where a single beam provides little discriminative margin, causing DT and SVR to exhibit large angular confusion ($\Delta\phi\!>\!75^\circ$) even without interference. Across both cases, SINR conditions push all markers further from the GT, with angle errors growing disproportionately relative to range errors, corroborating the CDF analysis. Fig.~\ref{fig:combined_side}(b) plots received power versus scan angle for GT $(19^\circ,\,0.770\,\mathrm{m})$. Under SNR, KNN and DT steer correctly to $19^\circ$, while SVR mispredicts by $26.5^\circ$, incurring a substantial power loss. Under SINR, RF degrades to $30.4^\circ$ and SVR to $47^\circ$, confirming that even a moderate angle MAE causes the beam to miss the UE entirely.
\subsection{Angle and Range Prediction Accuracy}
\label{subsec:cdfs}
Fig.~\ref{fig:cdf_angle} shows the CDF of absolute angle error $|\Delta\phi_r|$. Under the SNR baseline (dashed), KNN reaches the $90^{\mathrm{th}}$ percentile at only $0.90^\circ$; under SINR (solid), this grows to $3.45^\circ$ ($\Delta=+2.55^\circ$), while DT reaches $14.00^\circ$, the largest degradation among the four models. Fig.~\ref{fig:cdf_range} presents the range CDF, where the SNR--SINR gap is substantially narrower than in Fig.~\ref{fig:cdf_angle}. This contrast reveals a fundamental asymmetry: range is encoded in the \emph{absolute} power level of the fingerprint, which shifts uniformly under interference and is therefore relatively robust to angular beam leakage; angle, by contrast, is encoded in the \emph{relative} power ratios between beams, which are directly perturbed when an interferer lies at a beam-dominance boundary. KNN's $90^{\mathrm{th}}$-percentile range error grows by only $5.4$\,cm under interference, compared with a $2.55^\circ$ angle increase.
\subsection{Quantization Loss and Computational Complexity}
\label{subsec:quant}
Table~\ref{tab:quant_complexity}(A) reports mean 1-bit quantization loss per beam; Beam~$\phi_{b_1}$ incurs the highest average penalty (5.99--7.50\,dB) with worst-case $-18.31$\,dB at $d_2=0.4$\,m, while Beams~$\phi_{b_2}$--$\phi_{b_4}$ remain below 3.1\,dB. Despite this, binary phase control is sufficient for fingerprint-based localization since the regressors exploit relative inter-beam SNR ratios, which are preserved under a non-uniform quantization penalty. Table~\ref{tab:quant_complexity}(B) confirms all four models satisfy real-time constraints ($<\!1$\, ms inference): DT is lightest at $0.035$\,ms, while KNN achieves the best accuracy at only $0.199$\,ms, making it the recommended choice for online deployment.
\begin{table*}[!t]
\centering
\caption{\textbf{(A)} Mean quantization loss $\Delta P_r|$ (dB) for 1-bit vs.\ ideal phase
(positive = magnitude; \textit{Worst} row = signed minimum).
\textbf{(B)} Computational complexity ($N_{\rm tr}=480$, 5-run median).}
\label{tab:quant_complexity}
\renewcommand{\arraystretch}{1.05}
\begin{minipage}[t]{0.48\textwidth}
\centering
\footnotesize
\textbf{(A) Quantization Loss $\Delta P_r|$ (dB)}\\[2pt]
\setlength{\tabcolsep}{3.2pt}
\resizebox{\linewidth}{!}{%
\begin{tabular}{@{}lrrrrrr@{}}
\toprule
\textbf{Beam} & \multicolumn{6}{c}{$d_2$ (m)} \\
\cmidrule(lr){2-7}
& 0.2 & 0.4 & 0.6 & 0.8 & 1.0 & 1.2 \\
\midrule
$\phi_{b_1}^{\circ}$ & 5.99 & 6.30 & 6.89 & 7.19 & 7.37 & 7.50 \\
$\phi_{b_2}^{\circ}$ & 1.42 & 1.50 & 1.60 & 1.58 & 1.54 & 1.48 \\
$\phi_{b_3}^{\circ}$ & 2.45 & 3.06 & 2.94 & 2.78 & 2.65 & 2.58 \\
$\phi_{b_4}^{\circ}$ & 1.78 & 2.13 & 2.17 & 2.19 & 2.20 & 2.19 \\
\midrule
\textit{Worst} & $-13.55$ & $-18.31$ & $-15.71$ & $-13.06$ & $-12.26$ & $-13.00$ \\
\bottomrule
\end{tabular}%
}
\end{minipage}
\hfill
\begin{minipage}[t]{0.48\textwidth}
\centering
\footnotesize
\textbf{(B) Computational Complexity}\\[2pt]
\setlength{\tabcolsep}{3.0pt}
\resizebox{\linewidth}{!}{%
\begin{tabular}{@{}lccc@{}}
\toprule
\textbf{Model} & \textbf{Complexity} & \textbf{Train (ms)} & \textbf{Infer (ms)} \\
\midrule
KNN
& \begin{tabular}[c]{@{}c@{}}Train: $\mathcal{O}(1)$ \\ Infer: $\mathcal{O}(N_{\mathrm{tr}}B)$\end{tabular}
& 0.2 & 0.1986 \\

RF
& \begin{tabular}[c]{@{}c@{}}Train: $\mathcal{O}(TN_{\mathrm{tr}}B\log N_{\mathrm{tr}})$ \\ Infer: $\mathcal{O}(T\log N_{\mathrm{tr}})$\end{tabular}
& 64.2 & 2.3125 \\

DT
& \begin{tabular}[c]{@{}c@{}}Train: $\mathcal{O}(N_{\mathrm{tr}}B\log N_{\mathrm{tr}})$ \\ Infer: $\mathcal{O}(\log N_{\mathrm{tr}})$\end{tabular}
& 0.7 & 0.0353 \\

SVR
& \begin{tabular}[c]{@{}c@{}}Train: $\mathcal{O}(N_{\mathrm{tr}}^{2\text{--}3})$ \\ Infer: $\mathcal{O}(N_{\mathrm{SV}}B)$\end{tabular}
& 1.6 & 0.2355 \\
\bottomrule
\end{tabular}%
}
\end{minipage}
  \vspace{-4 mm}
\end{table*}
\section{Conclusion}
\label{Conclusion}
This paper investigated ML-based near-field UE localization in a blocked RIS-assisted mmWave system under cross-link interference. A four-beam swept fingerprint serves as the compact ML input, and an INR-constrained calibration strategy ensures a physically meaningful interference scenario. Among the regressors, KNN consistently achieved the lowest MAE of $0.37^\circ$ (angle) and $4$\,cm (range) under clean conditions, rising to $1.39^\circ$ and $7.6$\,cm under interference, alongside an inference latency of $0.19$\, ms per sample that satisfies real-time requirements. A fundamental asymmetry was identified, i.e., interference degraded the angle MAE by up to $4.4\times$ while the range MAE increased by at most $1.3\times$. This disparity arises because range is encoded in absolute beam power, which remains robust to angular leakage, whereas angle relies on inter-beam power ratios that are sensitive to user interference at a beam-dominance boundary.
Future work will extend the framework to distributed RIS-assisted deployments, to higher-bit codebooks such as 2-bit or 3-bit, to online adaptation of the interference coupling model, and to experimental validation of the proposed approach.

\bibliographystyle{IEEEtran}
\bibliography{biblio}

\end{document}